\documentstyle[multicol,aps,prl,epsf]{revtex}

\begin{document}
\draft

\title{Polaron versus bipolaron in conducting polymers:\\
a density matrix renormalization group study}

\author{Makoto Kuwabara\cite{post}, 
Yukihiro Shimoi, and Shuji Abe}
\address{Electrotechnical Laboratory, 
Umezono 1-1-4, Tsukuba, Ibaraki 305, Japan}
\date{\today}

\maketitle

\begin{abstract}
Competition between polaron and bipolaron in
conjugated polymers with nondegenerate ground state is
systematically studied in the extended Hubbard-Peierls model
with the symmetry-breaking Brazovskii-Kirova term,
using the density matrix renormalization
group method combined with lattice optimization in the adiabatic
approximation.
We demonstrate that the relative stability of a bipolaron over two
separated polarons sensitively depends on
both on-site Hubbard $U$ and nearest-neighbor repulsion $V$.
When $U$ is much larger than $V$, the bipolaron state is more
stabilized compared with mean field calculations.
\end{abstract}


\begin{multicols}{2}
\narrowtext

The formation of a bipolaron, in which two electrons are trapped in 
a self-induced
lattice deformation, is an important subject in many areas of condensed
matter physics, such as amorphous semiconductors\cite{Anderson}, insulating
crystals\cite{Hiramoto}, high temperature superconductors\cite{Nasu,Emin},
and low-dimensional materials\cite{RevModPhys}.
A simple model of localized linear electron-lattice coupling gives us a
criterion for bipolaron formation as that the electron-lattice coupling is
strong enough to surpass Coulomb repulsion between the two electrons ---
the situation being called "negative effective $U$"\cite{Anderson}.  
However, real situations are more delicate in the sense that bipolarons
possess internal degrees of freedom to allow modification of wavefunctions
and lattice deformations so as to reduce Coulomb repulsion and yet to
increase binding by lattice distortion. In addition, a purely electronic
mechanism of negative effective $U$ has been proposed\cite{Yoshida}.

In one-dimensional conjugated polymers, polarons and bipolarons belong to
the category of solitonic nonlinear excitations\cite{RevModPhys}.  
Solitons can exist in polymers with degenerate ground states, such as in
trans-polyacetylene, whereas polarons and bipolarons are candidates for
major charge carriers in polymers without ground state degeneracy, such as
in polythiophene and poly($p$-phyenylene vinylene).
Experimentally, the existence of bipolarons in doped conjugated polymers
has been controversial\cite{Furukawa-review,Conwell}.
The initial assignments of bipolarons relied upon the observation of two peaks
in doping-induced absorption spectra\cite{Ziemelis}.  
However, theoretical calculations\cite{Fesser,Shimoi_DIA} as well as
experimental studies of doped oligomers\cite{Furukawa-review,Fesser} have
established that the two-peak feature actually corresponds to a polaron
rather than a bipolaron.
On the other hand, the existence of bipolarons has been suggested by 
the observation of a single 
photoinduced absorption peak in an improved sample of
polythiophene\cite{Vardeny}.

On the theoretical side, models with neglecting electron-electron (e-e)
interactions predict that
a bipolaron (BP) is always energetically more stable than
a separated pair of polarons (2P)\cite{Baeriswyl}.  
The effect of long-range e-e interaction was studied systematically within
the unrestricted Hartree-Fock (UHF) approximation\cite{Shimoi}, and it was
shown that the interaction significantly suppresses the stability of BP.
A recent study in the extended Hubbard-Peierls model by using 
the density matrix renormalization group (DMRG) method\cite{White} has
shown that BP is
favored over 2P even for a very large on-site repulsion $U$\cite{Su}.
However, off-site interactions were not taken into account in that study,
and, in view of the spatial extensions of a polaron and a bipolaron, it
is very important to take those interactions as well.

In the present paper, we investigate this problem systematically by using
the DMRG method and by taking both on-site $U$ and nearest neighbor 
repulsion $V$ into account.
The DMRG method developed originally by White\cite{White} for quantum spin
systems is 
quite a powerful method to treat strongly correlated electron systems in
one dimension.
For the present purpose we combine the DMRG method with lattice
optimization by using the Hellmann-Feynman force equilibrium condition.
We will demonstrate that higher order correlation effects beyond the mean
field approximation are crucial for the formation of BP and
that the stability of BP is
very sensitive to both $U$ and $V$.

The model Hamiltonian is the one-dimensional extended Hubbard-Peierls
Hamiltonian,
\begin{eqnarray}
H & = & -\sum_{i,s} t_{i,i+1} [c^\dagger_{i,s} c_{i+1,s} + h.c.] \nonumber \\
  &   & +U\sum_i n_{i\uparrow} n_{i\downarrow}
        +V\sum_i (n_i -1)(n_{i+1}-1) \nonumber \\
  &   & +\frac{K}{2} \sum_i y_i^2
        +\Gamma \sum_i y_i.
\label{eq:PHmodel}
\end{eqnarray}
Here, the creation operator of a $\pi$ electron 
with spin $s$ at site $i$ is denoted by $c^\dagger_{i,s}$, and 
the number operators are defined as
$n_{i,s}=c^\dagger_{i,s}c_{i,s}$ and $n_i=\sum_s n_{i,s}$.  
The transfer integral, $t_{i,i+1}$, is assumed in the form
\begin{equation}
t_{i,i+1} = [1-(-1)^i\delta_0]t + \alpha y_i,
\end{equation}
where $\delta_0$ is the Brazovskii-Kirova symmetry-breaking 
parameter to lift the degeneracy of the ground state\cite{BK},
$y_i$ the change of bond length between $i$-th and $(i+1)$-th sites, 
and $\alpha$ the electron-lattice coupling constant of 
the Su-Schrieffer-Heeger type\cite{SSH}.  
The parameter $K$ is the spring constant for the change of bond length, 
and the last term in 
eq. (\ref{eq:PHmodel}) is added to keep the chain length constant.  
We take the dimensionless electron-lattice coupling 
constant $\lambda=2\alpha^2/\pi tK=0.136$ and $\alpha/K=0.084$\AA \ 
as in Ref. \cite{Shimoi}.  
Although the relative stability of BP and 2P depends on the
electron-lattice coupling, we fix this parameter and vary e-e interaction
strengths $U$ and $V$, for clarity.

We apply the finite-system DMRG algorithm\cite{White} to treat 
the electronic ground state. 
The open boundary condition is imposed to the system with $N$ sites.  
The lattice is treated within the adiabatic approximation.  
The bond variable $y_i$ is determined by using the Hellmann-Feynman 
force equilibrium condition\cite{Shimoi,Su},
\begin{equation}
y_i = -\frac{1}{K} 
     \left( 2\alpha p_{i,i+1} +\Gamma \right), \label{eq:y_i}
\end{equation}
where $p_{i,i+1}=\sum_s < c_{i,s}^\dagger c_{i+1,s} >$ with 
$<\cdots>$ denoting the ground state expectation value.  
The condition $\sum y_i = 0$ gives 
$
\Gamma = - (2\alpha/N-1) \sum_m p_{m,m+1}.
$
Thus the lattice configuration $\{y_i\}$ and the electronic ground state 
must be determined in a self-consistent fashion.
We use an iterative method as follows: 
(i) Set an initial lattice configuration $\{ y_i^{(0)} \}$.  
(ii) Calculate the electronic ground state by the infinite-system 
DMRG method for $\{ y_i^{(0)} \}$.  
(iii) Carry out the finite-size DMRG procedure to refine the ground state.
(iv) Calculate new lattice configuration $\{ y_i^{(1)} \}$ 
from eq. (\ref{eq:y_i}).  
(v) Replace $\{ y_i^{(0)} \}$ with $\{ y_i^{(1)} \}$ and go to step (ii).  
The procedure is continued until the difference between 
$\{ y_i^{(0)} \}$ and $\{ y_i^{(1)} \}$ becomes negligibly small.  
In the present DMRG calculations, we truncate the Hilbert space by keeping 
$m=80$ states in each block, which turned out to be sufficient to achieve 
negligible $m$-dependence of calculated results. 
We have also checked the accuracy of our DMRG calculations by comparing 
with the results of numerical exact diagonalization up to 12 sites, 
with the exact solution of the Hubbard model, and 
with exact numerical results in the non-interacting case ($U=V=0$).

We consider the doping of two electrons in a finite chain to examine the 
formation of BP or 2P.  
We present results for the system size $N=120$ and 
the total electron number $N_{\rm e}=122$.  
Here electrons with up and down spins are added 
to the half-filled system 
so as to give $S_{\rm z}=0$.  
To obtain a 2P solution, we can equally use 
two up spin electrons with $S_{\rm z}=1$ instead,
because
the interference between the two well-separated polarons is negligible.  
This saves a lot of 
CPU time especially when 2P is less stable than BP.  
We have extended calculations to larger systems
(up to $N \sim 200$) and confirmed that the results are not sensitive to the
system size.

\begin{figure}
\epsfxsize=3 in\epsfbox{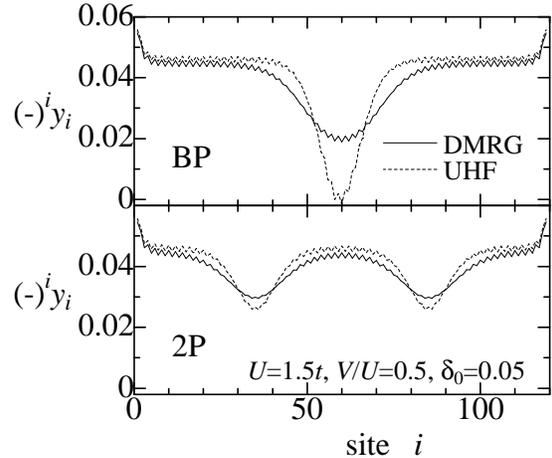}
\caption{
Optimized bond alternation patterns of a bipolaron and a pair of polarons 
calculated in the DMRG (solid line) and UHF (broken line) methods
for $U=1.5t$, $V/U=0.5$, and $\delta_0=0.05$.
}
\label{fig1}
\end{figure}

In Fig. 1 we show an example of obtained lattice distortions of BP and 2P
along with corresponding UHF results for comparison.
In this example with $U=1.5t$ and $V/U=0.5$, we see that
polaronic distortions (especially for a bipolaron) are wider and shallower
in the DMRG than in the UHF.
However, if we go to much smaller $V/U$, this tendency is reversed, i.e.,
the distortion becomes narrower and deeper in the DMRG.
This implies that higher order correlation beyond the UHF
works in opposite directions with respect to $U$ and $V$.

Figure 2 displays the energy difference $\Delta E$ between
the obtained BP and 2P states
as a function of $U$ with constant $V/U$ for the case $\delta_0=0.05$.
$\Delta E>0$ implies that BP is more stable than 2P.
The UHF results are also depicted for comparison.
In the case of $V/U=0.5$, BP is more stable than 2P for $U$ below
a critical value $U_{\rm c}\sim 1.8t$.  
The allover behavior is qualitatively similar to
the UHF result, although $U_{\rm c}$ is substantially larger.
We could not find out a metastable BP state for
$U>U_{\rm c}$
in the DMRG calculation in contrast to the UHF result.
In this case the iteration procedure always leads to a 2P state
with $S_{\rm z}=0$ even if we start from a BP configuration.
This is the reason why the DMRG curve for $V/U=0.5$ in Fig. 2 is terminated
at $U_{\rm c}$ above which the 2P state is the only self-consistent solution.

\begin{figure}
\epsfxsize=3 in\epsfbox{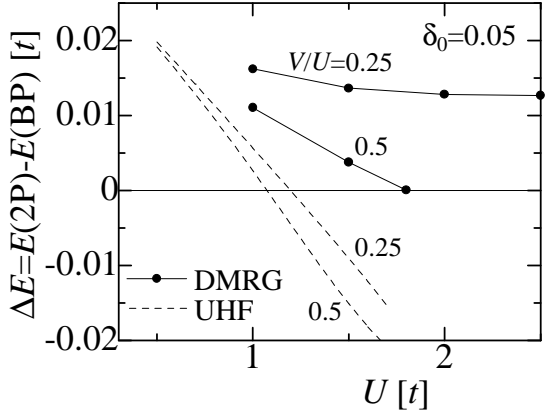}
\caption{
Energy difference between the bipolaron and two-polaron states against $U$
calculated by the DMRG method (solid lines) and the UHF approximation 
(broken lines) for $V/U=0.5$ and $0.25$ with $\delta_0$=0.05.
}
\label{fig2}
\end{figure}

On the other hand,
$\Delta E$ for $V/U=0.25$ in Fig. 2 is always positive
and does not show any indication of crossing $\Delta E=0$
even for much larger $U$.
That is, BP is always stable in this case.
This is completely different from the UHF result, which
is not very sensitive to $V/U$.
We note that this result is consistent with Ref.\cite{Su} for
the limiting case of $V=0$.

The destabilization of the BP state with increasing $U$ is understandable
at the UHF level as follows\cite{Shimoi}.
In restricting our consideration to the effects of $U$,
a bipolaron with spatial extension
$\xi_{\rm BP}$ costs energy $\sim U/\xi_{\rm BP}$, while a polaron
with an extension $\xi_{\rm P}$ hardly costs energy because
of largely spin-polarized excess charges in the case of large $U$.
Higher order correlation may reduce the probability of double occupancy,
thus weakening the $U$-dependence of $\Delta E$ in the DMRG method.

We have carried out similar calculations for various $\delta_0$.
By plotting the critical value $U_{\rm c}$ for BP destabilization
against $\delta_0$, we obtain the ground state phase diagram
in the parameter space of $\delta_0$ and $U$.
Figure 3 is the obtained phase diagram for $V/U=0.5$, where
the solid (broken) line indicates the phase boundary between the regions
of stable BP and 2P in the DMRG (UHF) results.
The phase boundary becomes lower with raising $\delta_0$, because
the increment of $\delta_0$ results in the decrement of $\xi_{\rm BP}$ and
$\xi_{\rm P}$, thereby enhancing the effect of Coulomb repulsion.
Higher order correlation stabilizes the BP state, as mentioned above.
In the case of $V/U=0.25$
we did not find out the region of 2P being more stable than BP
for any $\delta_0$ in the DMRG method.

\begin{figure}
\epsfxsize=3 in \epsfbox{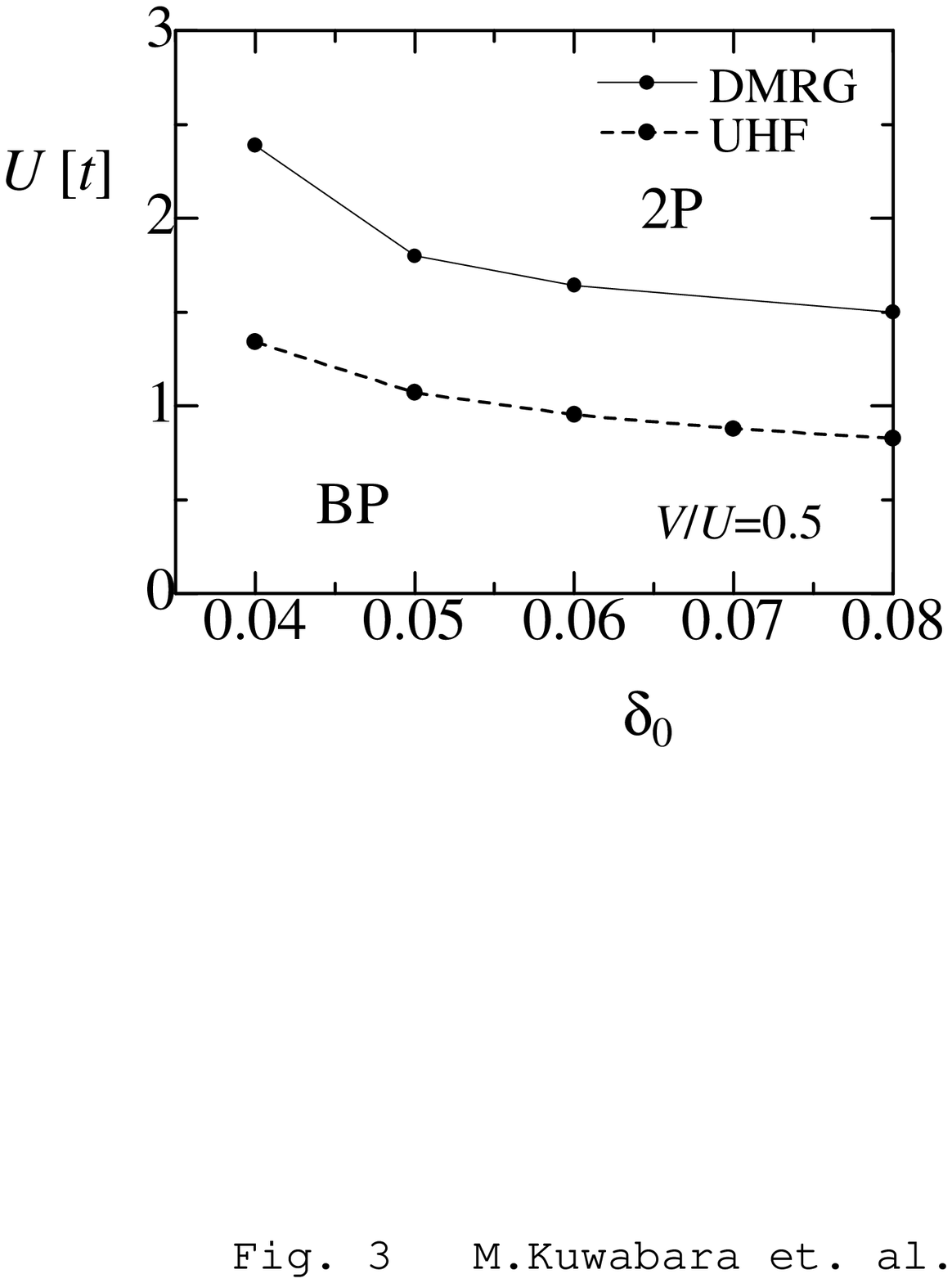}
\caption{
Ground state phase diagram of the two-electron doped system with $V/U=0.5$.
The solid (broken) line indicates the phase boundary between the regions of
a bipolaron (BP) and two separated polarons (2P) obtained by the DMRG method
(the UHF approximation).
}
\label{fig3}
\end{figure}

In order to clarify the role of higher order correlation beyond the
UHF approximation for the energies of 2P and BP,
we have evaluated the fluctuation energies in the obtained DMRG wave
functions.
The spin fluctuation energy, 
$U \sum_i <(n_{i\uparrow}-<n_{i\uparrow}>)
           (n_{i\downarrow}-<n_{i\downarrow}>)>$,
is always larger for 2P than in BP,
destabilizing the 2P state.
Its difference between 2P and BP grows with increasing $U$,
so that this contribution is more important for smaller $V/U$.
On the other hand, the charge fluctuation energy,
$V \sum_{i,s,s'} <(n_{i,s}-<n_{i,s}>)(n_{i+1,s'}-<n_{i+1,s'}>)>$,
is almost the same for 2P and BP,
so that the net effect to the relative stability is negligible.
Therefore, the main difference between the DMRG and
UHF results comes from spin fluctuations
associated with on-site Coulomb repulsion $U$.

In summary, we have investigated the effect of e-e interactions
on the spatial profiles and energies of polarons and bipolarons
in conjugated polymers by using the DMRG method.
We have demonstrated that the relative stability of a bipolaron over a pair of
separated polarons is sensitive to both on-site $U$ and nearest neighbor $V$.
Spin fluctuations associated with on-site $U$
contributes to the stabilization of a bipolaron compared with the results of
the UHF approximation.
On the other hand, nearest neighbor $V$ works in the direction of
dissociating a bipolaron into two polarons in the same way as the 
UHF calculation.

The present study is limited to the extended Hubbard-Peierls model with
short range interactions, while longer range interactions such as the
Ohno potential should be taken into account to discuss situations
in real materials.
Judging from the roles of $U$ and $V$ clarified in the present work,
we expect that the inclusion of longer range interactions would
not cause qualitative differences but enhance the tendency of
stabilizing polarons by $V$.

The authors are grateful to Prof. S. Ramasesha for clear
lectures on the DMRG method, and to Prof. Y. Furukawa
for stimulating discussions on experimental results.
M.K. would like to thank Dr. T. Yanagisawa for
helpful advice about numerical diagonalization methods.
Numerical calculations were performed at the
Research Information Processing System Center,
Agency of Industrial Science and Technology.

\def\PRB#1{Phys. Rev. B {\bf #1}}
\def\PRL#1{Phys. Rev. Lett. {\bf #1}}
\def\SSC{Solid State Commun. }
\def\JPS#1{J. Phys. Soc. Jpn. {\bf #1}}
\def\SM#1{Synth. Met. {\bf #1}}
\def\PL{Phys. Lett. }

\end{multicols}{2}
\end{document}